\documentclass[a4paper, oneside, twocolumn]{scrartcl}

\usepackage[english]{babel}
\usepackage{listings}
\usepackage{graphicx}
\usepackage{color}
\usepackage{colortbl}
\usepackage[bookmarks,raiselinks,pageanchor,hyperindex]{hyperref}
\usepackage{amsmath,amssymb,amsfonts,amstext,bbm}
\usepackage{geometry}
\usepackage[normalem]{ulem}
\usepackage{xspace}
\usepackage[latin1]{inputenc}
\usepackage[style=ieee]{biblatex}
\usepackage{csquotes}
\usepackage{setspace}
\usepackage[justification=centering,singlelinecheck=false,labelfont={bf,footnotesize,sf},font={footnotesize,sf}, aboveskip=1em,belowskip=-1em]{caption}
\usepackage{pgf}

\usepackage{glossaries}
\usepackage{todonotes}
\usepackage{booktabs}
\usepackage{tikz}
\usepackage[binary-units, per-mode=symbol-or-fraction]{siunitx}
\usepackage{nicefrac}
\usepackage[capitalise]{cleveref}
\usepackage[printonlyused]{acronym}
\glsdisablehyper

\DeclareMathOperator*{\argmax}{argmax}

\definecolor{color0}{HTML}{1f77b4}  
\definecolor{color1}{HTML}{ff7f0e}  
\definecolor{color2}{HTML}{2ca02c}  
\definecolor{color3}{HTML}{d62728}  
\usepackage{changes}

\newacronym{adc}{ADC}{analog-to-digital converter}
\newacronym{anncore}{ANNCORE}{analog neural network core}
\newacronym{ann}{ANN}{artificial neural network}
\newacronym{awgn}{AWGN}{additive white Gaussian noise}
\newacronym{asic}{ASIC}{application-specific integrated circuit}
\newacronym{bss-2}{BSS-2}{BrainScaleS-2}
\newacronym{ber}{BER}{bit error rate}
\newacronym{bptt}{BPTT}{backpropagation through time}
\newacronym{cadc}{CADC}{columnar \acrshort{adc}}
\newacronym{cd}{CD}{chromatic dispersion}
\newacronym{dsp}{DSP}{digital signal processing}
\newacronym{fec}{FEC}{forward error correction}
\newacronym{li}{LI}{leaky-integrate}
\newacronym{itl}{ITL}{in-the-loop}
\newacronym{imdd}{IM/DD}{intensity-modulation / direct-detection}
\newacronym{lif}{LIF}{leaky-integrate and fire}
\newacronym{lmmse}{LMMSE}{linear minimum mean square error}
\newacronym{pd}{PD}{photodiode}
\newacronym{nn}{NN}{neural network}
\newacronym{snn}{SNN}{spiking neural network}
\newacronym{ode}{ODE}{ordinary differential equation}
\newacronym{pam4}{PAM4}{pulse amplitude modulation 4-level}
\newacronym{rrc}{RRC}{root-raised-cosine}

\begin{document}
\selectlanguage{english}

\title{Spiking Neural Network Equalization on Neuromorphic Hardware for IM/DD Optical Communication}

\author{
	Elias Arnold\thanks{Electronic Vision(s), Kirchhoff-Institute for Physics, Heidelberg University, Germany}
	\and
	Georg B\"ocherer\thanks{Huawei Technologies Duesseldorf GmbH, Munich Research Center, Germany}
	\and
	Eric M\"uller\footnotemark[1]
	\and
	Philipp Spilger\footnotemark[1]
	\and
	Johannes Schemmel\footnotemark[1]
	\and
	Stefano Calabr\`o\footnotemark[2]
	\and
	Maxim Kuschnerov\footnotemark[2]
}
\date{2022-05-11}

\maketitle

\begin{abstract}
A spiking neural network (SNN) nonlinear equalizer model is implemented on the mixed-signal neuromorphic hardware system BrainScaleS-2 and evaluated for an IM/DD link. The BER 2e-3 is achieved with a hardware penalty less than \SI{1}{dB}, outperforming numeric linear equalization.
\end{abstract}
\glsresetall

\section{Introduction}

\begin{figure}[htb]
    \centering
    \includegraphics[width=1.0\linewidth]{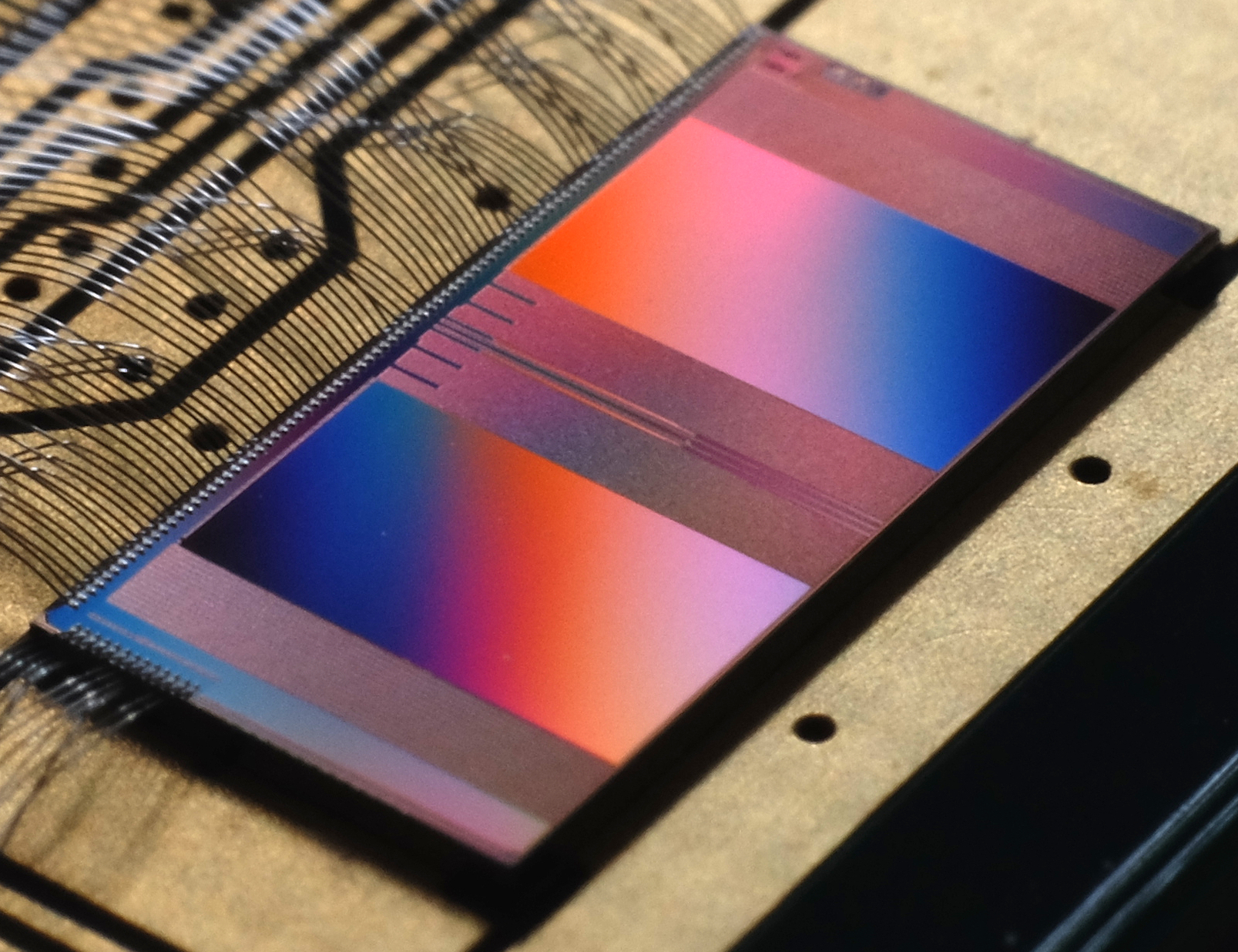}%
    \caption{%
        A \acrlong{bss-2} \gls{asic} bonded to a carrier board. The chip is about \SI{4}{\milli\meter}~\texttimes~\SI{8}{\milli\meter} in size.
    }
    \label{fig:bss2}
\end{figure}

Cloud services cause an exponentially growing traffic in data centers. This requires optical transceivers to operate at lower power and lower cost. Because \gls{dsp} has high power consumption, recent research envisions replacing the \gls{dsp} in part with an analog frontend with lower power consumption. One approach is photonic neuromorphic computing~\cite{shastri2021photonics}, which has been proposed, e.g., for \gls{cd} compensation and nonlinear equalization in short-reach optical transmission~\cite{li2021micro,ranzini2021experimental}. A second approach is analog electronic neuromorphic computing, which implements \glspl{snn}~\cite{gerstner2014neuronal} in analog hardware~\cite{pehle2022brainscales2}, adopting the brain's unique power efficiency by imitating the basic functioning of the human brain. Recently, \gls{itl} training of \glspl{snn} on analog hardware~\cite{schmitt2017neuromorphic} has shown promising results by achieving state-of-the-art performance in inference tasks~\cite{cramer2022surrogate}.
Although photonics is operating faster than electronic hardware, the latter scales better and can therefore achieve higher throughput by parallelization. Electronic hardware is therefore tailored for low power signal processing.

In \textcite{arnold2022spiking},
we design and evaluate an \gls{snn} equalizer in a software simulation for the detection of a \gls{pam4} signal for an \gls{imdd} link, impaired by \gls{cd} and \gls{awgn}. The results show that in principle, \glspl{snn} can perform as well as nonlinear \gls{ann} equalizers, outperforming linear equalizers.

However, to assess the applicability of \glspl{snn} for equalization in practical systems, evaluation in hardware is essential. 

In this work, we design and implement \glspl{snn} for joint equalization and demapping on the mixed-signal neuromorphic \gls{bss-2} system~\cite{pehle2022brainscales2} displayed in Fig.~\ref{fig:bss2}. We showcase that an \gls{snn} equalizer/demapper trained on the analog substrate efficiently detects a \gls{pam4} signal of an \gls{imdd} link
enabling a \SI{200}{\giga\bit\per\second} transmission with \SI{12}{\percent} overhead hard\added{-}decision \gls{fec}, outperforming numeric linear equalization. Furthermore, we compare the \gls{ber} achieved on \gls{bss-2} to \gls{ann} and \gls{snn} equalizers simulated in software.

\section{Simulated \acrshort{imdd} Link}

\begin{figure*}[tbh]
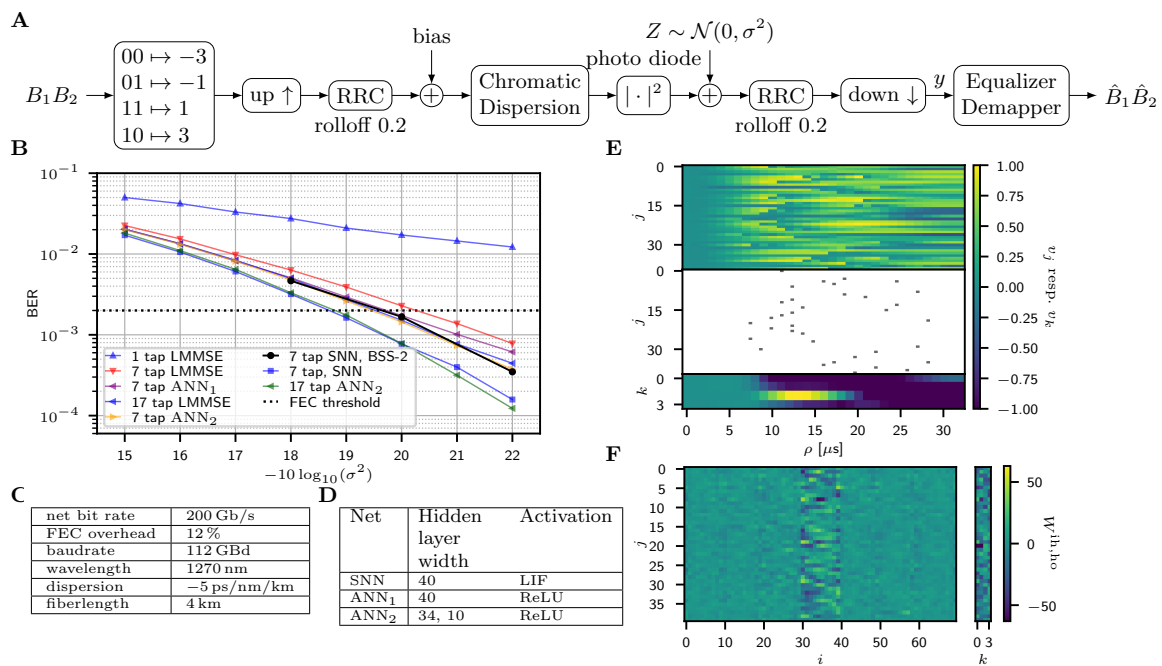

    \centering
    \footnotesize
    \tikzset{
        panel/.style={
            inner sep=0pt, outer sep=0, execute at begin node={\tikzset{anchor=center, inner sep=.33333em}}},
        label/.style={anchor=north west, inner sep=0, outer sep=0}}
	\scalebox{0.9}{%
	\begin{tikzpicture}

    \node[panel, anchor=north west] (a) at (0,  4) {%
        \begin{tikzpicture}
    \tikzset{device/.style={align=center, draw, rounded corners}}
    \matrix[column sep=0.4cm, ampersand replacement=\&]{
    \node(src){$B_1B_2$};
    \& \node[device,align=left](map){$00\mapsto -3$\\
    $01\mapsto -1$\\
    $11\mapsto 1$\\
    $10\mapsto 3$};
    \& \node[device](up){up $\uparrow$};
    \& \node[device](rrc){RRC};
    \& \node[circle, draw, inner sep=0cm](bias){$+$};
    \& \node[device](cd){Chromatic\\Dispersion};
    \& \node[device](pd){$|\cdot|^2$};
    \& \node[circle, draw, inner sep=0cm](noise){$+$};
    \& \node[device](rrc_rx){RRC};
    \& \node[device](down){down $\downarrow$};
    \& \node[device](demap){Equalizer\\ Demapper};
    \& \node(sink){$\hat{B}_1\hat{B}_2$};\\
    };
    \draw[-latex](src)--(map);
    \draw[-latex](map)--(up);
    \draw[-latex](up)--(rrc);
    \draw[-latex](rrc)--(bias);
    \draw[-latex](bias)--(cd);
    \draw[-latex](cd)--(pd);
    \draw[-latex](pd)--(noise);
    \draw[-latex](noise)--(rrc_rx);
    \draw[-latex](rrc_rx)--(down);
    \draw[-latex](down)--node[above]{$y$}(demap);
    \draw[-latex](demap)--(sink);
    
    \node[above=0.5cm of bias](bias_val){bias};
    \draw[-latex](bias_val)--(bias);
    
    \node[above=0.5cm of noise](noise_val){$Z\sim\mathcal{N}(0,\sigma^2)$};
    \draw[-latex](noise_val)--(noise);
    
    \node[below=0cm of rrc]{rolloff $0.2$};
    \node[below=0cm of rrc_rx]{rolloff $0.2$};
    
    \node[above=0cm of pd]{photo diode};
\end{tikzpicture}};%
    \node[label] at (a.north west) {\textbf{A}};
    
	\node[panel, anchor=north west] (c) at (0.3, -3.3) {%
        \tiny
\begin{tabular}{|l|l|}\hline
    net bit rate & \SI{200}{Gb/s} \\\hline
    FEC overhead & \SI{12}{\%} \\\hline
    baudrate & \SI{112}{GBd} \\\hline
    wavelength & \SI{1270}{nm} \\\hline
    dispersion & \SI{-5}{ps/nm/km} \\\hline
    fiberlength & \SI{4}{km}\\\hline
\end{tabular}
};%
    \node[label] at (0., -3.0) {\textbf{C}};
    
    \node[panel, anchor=north west] (e) at (9., 1.9) {%
        \input{hw_traces.pgf}};%
    \node[label] at (8.7, 2.1) {\textbf{E}};
    
    \node[panel, anchor=north west] (f) at (9., -2.6) {%
        \input{weight_matrices.pgf}};%
    \node[label] at (8.7, -2.4) {\textbf{F}};
    
    \node[panel, anchor=north west] (b) at (0.2, 1.8) {%
        \input{snr_sweep.pgf}};%
    \node[label] at (0, 2.1) {\textbf{B}};%
    %
    \node[panel, anchor=north west] (d) at (4.70, -3.3) {%
        \setlength{\tabcolsep}{4pt}
\tiny
\begin{tabular}{|l|p{1.2cm}l|}\hline
    \scriptsize Net & \scriptsize Hidden layer width & \scriptsize Activation \\ \hline
    SNN & 40 & LIF \\\hline
    ANN$_1$ & 40 & ReLU \\\hline
    ANN$_2$ & 34, 10 & ReLU \\\hline
\end{tabular}
};%
    \node[label] at (4.5, -3) {\textbf{D}};
    
    \end{tikzpicture}
	}
    \caption{\textbf{(A)} Simulated \acrshort{imdd} link. Schematic taken from \textcite{arnold2022spiking}. \textbf{(B)} Comparison of \acrshort{ber} results for transmission over a simulated \acrshort{imdd} link with PAM4 constellations. \textbf{(C)} Considered \acrshort{imdd} parameters. \textbf{(D)} \Acrshort{nn} parameters of demappers used. \textbf{(E)} Experiment observables of a single run recorded on \acrshort{bss-2}. \textbf{(F)} Weight matrices learned on \acrshort{bss-2}.}
    \label{fig:models}
\end{figure*}

As an application scenario, we simulate a \SI{200}{Gb/s} \gls{imdd} transmission over \SI{4}{\km} in the O-band. We assume a \SI{12}{\percent} overhead \gls{fec} with a \gls{ber} threshold \num{2e-3} and a corresponding baudrate of \SI{112}{GBd}. Our model is displayed in \cref{fig:models}A and parameters are summarized in \cref{fig:models}C.
A bit sequence $([B_1B_2]^t)_{t\in\mathbb{N}}$ is mapped to a \gls{pam4} signal, which is upsampled and filtered with a \gls{rrc}, after which a positive bias is added. Transmission through the fiber is then simulated by applying \gls{cd}. 
At the receiver, a \gls{pd} squares the signal and \gls{awgn} is added. The signal is then \gls{rrc} filtered and downsampled. The resulting sequence $(y^t)_{t\in\mathbb{N}}$ is equalized and demapped, resulting in the bit decisions $([\hat{B}_1\hat{B}_2]^t)_{t\in\mathbb{N}}$.

\section{Spiking Neural Network Equalizer/Demapper}
Besides minor adaptations, we consider an \gls{snn} equalizer/demapper as described in our recent work~\cite{arnold2022spiking} and which we summarize next. Our \gls{snn} architecture has one hidden layer of 40 \gls{lif} neurons~\cite[Sec.~1.3]{gerstner2014neuronal}, projecting spike events onto four \gls{li} readout neurons~\cite[Sec.~1.3]{gerstner2014neuronal}.
The membrane potential $v_j$ of \Gls{lif} neurons is described by the \gls{ode}
\begin{align}
    \tau_\text{m} \dot{v}_j(\rho) = - \left( v_j(\rho) - v_\text{l}\right) + I_j(\rho),
\end{align}
with $v_\text{l}$ the leak potential, $\tau_\text{m}$ the membrane time constant and $I_j$ the current onto neuron $j$ caused by pre-synaptic events.
As $v_j$ exceeds a threshold $\vartheta$, neuron $j$ sends out a spike at time $\rho^s_j$ and is set to a reset potential $v_\text{r}$.
The current $I_{j}$ onto neuron $j$ is given by the exponentially filtered pre-synaptic spike train,
\begin{align}
    I_{j}(\rho) = \sum_i \sum_{\lbrace \rho^s_i \rbrace} W_{ji} \Theta \left( \rho-\rho^s_i \right)\exp\left(\frac{\rho^s_i-\rho}{\tau_\text{syn}}\right)
\end{align}
with synaptic time constant $\tau_\text{syn}$ and $\rho^s_i$ the spike times of the pre-synaptic neurons $\lbrace i \rbrace_{i=1}^{< n^\text{i}}$.
The membrane potential $v_k$ of \gls{li} neurons is subject to the same dynamics, without the ability to spike.
That is, current $I_k$, caused by spikes of the hidden \gls{lif} neurons $\lbrace j \rbrace_{j=0}^{<n^\text{h}}$, is integrated onto the \gls{li} membranes.

For demapping a transmitted sample $y^t$, we assign each sample $y^t_l \in \mathcal{C} = \left[ y^{t- \lfloor \nicefrac{n_{\text{tap}}}{2}\rfloor}, y^{t+\lfloor \nicefrac{n_{\text{tap}}}{2}\rfloor}\right]$ ($n_\text{tap}$ odd and $l\in\mathbb{N}^{<n_\text{tap}}_0$ indexing $\mathcal{C}$) $n^\text{i, tap} = 10$ input neurons, emitting spikes at times $\rho^s_i$ with $i = n^\text{i, tap}\cdot l + h$ and $h \in \mathbb{N}_0^{<n^\text{i, tap}}$.
Note that the time $\rho$ is the time axis within the time frame between two samples at time steps $t$ and $t+1$.
The spike times $\rho^s_i$ are given by a linear scaling of the distances of $y^t_l$ to reference points $\chi_h$. 
Finally, $y^t$ is labeled with the class $k \in \lbrace 0, 1, 2, 3\rbrace$ for which the corresponding readout neuron has the maximum membrane value over time, i.e.\ $\argmax_k\max_\rho v_k(\rho)$.
Thus, the network's objective is to learn to adjust the hidden neurons such that their spikes tune the membrane potentials of the output neurons meaningfully.

Because of the non-differentiable output of spiking \gls{lif} neurons, we use surrogate gradients (SuperSpike~\cite{neftci2019superspike}) in conjunction with \gls{bptt} to train our \glspl{snn}. Weights are optimized with the Adam optimizer~\cite{kingma2012adam}.

\section{BrainScaleS-2 System}
We deploy our \gls{snn} equalizer/demapper on the accelerated mixed-signal neuromorphic \gls{bss-2} system, developed at Heidelberg University~\cite{pehle2022brainscales2} in Germany.
A photo of the chip is shown in \cref{fig:bss2}.
On its \acrlong{anncore}, it emulates up to \num{512} \gls{lif} neurons and \SI{128}{k} synapses in analog circuits in parallel and in continuous time.
Hardware synapses have 6-bit weights, which can be configured as inhibitory (negative sign) or excitatory (positive sign).
The neurons communicate via digital spike events.
Each neuron is parameterized individually to exhibit the desired dynamic.
\gls{li} neurons can be realized by disabling the spiking mechanism.
To facilitate training, \gls{bss-2} provides a \gls{cadc} allowing to read out neuron membrane voltages in parallel. Effectively, this allows \gls{itl} learning with surrogate gradients~\cite{cramer2022surrogate} where the forward pass is performed on-chip and weight updates are computed on the host computer~\cite{schmitt2017neuromorphic}.

For training our \glspl{snn} on the \gls{bss-2} system we utilize the \texttt{PyTorch}-based~\cite{paszke2019pytorch} software framework \texttt{hxtorch.snn} supporting network execution on hardware and in simulation~\cite{mueller2022scalable}.
The \gls{bss-2} software stack translates the high-level experiment description into a corresponding hardware configuration including stimulus data, uploads the experiment to an FPGA-based real-time experiment controller, and post-processes recorded output data.

Processing one sample $y^t$ by the \gls{bss-2} takes \SI{30}{\micro\second}. In practical implementations, this can be speeded up, furthermore, many samples can be processed in parallel for achieving the required throughput.

\section{Results}
In \cref{fig:models}B, we plot the \glspl{ber} achieved with the 7-tap \gls{snn} equalizer/demapper, trained in software and on the analog \gls{bss-2} system. For comparison, we also plot the \glspl{ber} of software \gls{lmmse} equalizers followed by a demapper with \gls{ber}-optimized decision boundaries.
As further references, we train software \gls{ann} equalizer/demappers with one and two hidden layers, see~\cref{fig:models}D. We observe that the simulated \gls{snn} performs better than the \glspl{ann}.
The results show that the 7-tap \gls{snn} equalizer/demapper trained on \gls{bss-2} outperforms the 7-tap \gls{lmmse}.
For a \gls{ber} of \num{2e-3}, we observe a hardware penalty of less than \SI{1}{\dB} between the simulated \glspl{snn} and the \gls{bss-2} \gls{snn}.

\Cref{fig:models}E exemplifies the evolution of the hidden \gls{lif} neurons' analog membrane potentials $v_j$ (upper) and their spikes (center) together with the readout traces $v_k$ (lower) along the time $\rho$ on the \gls{bss-2} system.
The hidden spikes push the membrane of the correct readout neuron $k=2$ upwards while suppressing the traces of the other neurons $k=0,1,3$, indicating a confident decision.
The corresponding weight matrices $W^\text{ih}$ and $W^\text{ho}$ are depicted in \cref{fig:models}F.
The weights of the input neurons, receiving events from the central tap, are dominating since they encode the sample to classify, $y^t$.

\section{Conclusions}

In this work, we have implemented an \gls{snn} on the mixed-signal neuromorphic hardware system BrainScaleS-2 (\gls{bss-2}). While we observed a hardware penalty slightly below \SI{1}{\dB} at a \gls{ber} of \num{2e-3}, the \gls{bss-2} \gls{snn} outperforms a simulated linear equalizer with the same number of taps, thanks to nonlinear processing. 

The presented results confirm that neuromorphic hardware can provide the reliability required by signal processing in optical transceivers. Promising directions for future research include reducing the hardware penalty, reducing the architectural complexity and increasing the intrinsic speed of the \gls{snn}, and optimizing the input and output interfaces of the \gls{snn}. Furthermore, processing by the \gls{snn} must be parallelized. Also, the effective power consumption of neuromorphic signal processing should be analyzed and compared to digital processing.

In a future hardware implementation, spikes could be generated directly from the electrical signal coming from the photo diode, thereby replacing the power-hungry \gls{adc}. In this work, we have considered hard decision demapping. Another interesting direction is the design of an \gls{snn} equalizer/demapper with soft output.

\section{Acknowledgements}
We thank
S.\ Billaudelle,
L.\ Blessing,
B.\ Cramer,
and
C.\ Pehle
for fruitful discussions,
J.\ Weis
for basic 
hardware parameterization,
C.\ Mauch
for keeping the \gls{bss-2} system well-oiled,
and
all members of the Electronic Vision(s) research group who contributed to the \gls{bss-2} system.

\section*{Funding}
The contributions of the Electronic Vision(s) group\footnotemark[1] have been supported by
the EC Horizon 2020 Framework Programme under grant agreements
785907 (HBP SGA2)
and
945539 (HBP SGA3),
the Deutsche Forschungsgemeinschaft (DFG, German Research Foundation) under Germany's Excellence Strategy EXC 2181/1-390900948 (the Heidelberg STRUCTURES Excellence Cluster),
the Helmholtz Association Initiative and Networking Fund [Advanced Computing Architectures (ACA)] under Project SO-092.

\printbibliography

\end{document}